# Digital interfaces of historical newspapers: opportunities, restrictions and recommendations

Eva Pfanzelter[1]*, Sarah Oberbichler[1], Jani Marjanen[2], Pierre-Carl Langlais[3], Stefan Hechl[1]

[1]University of Innsbruck, Austria
[2]Universtity of Helsinki, Finland
[3]University Paul-Valéry Montpellier, France
*Corresponding author: Eva Pfanzelter (Eva.Pfanzelter@uibk.ac.at)

**Abstract**
Many libraries offer free access to digitised historical newspapers via user interfaces. After an initial period of search and filter options as the only features, the availability of more advanced tools and the desire for more options among users has ushered in a period of interface development. However, this raises a number of open questions and challenges. For example, how can we provide interfaces for different user groups? What tools should be available on interfaces and how can we avoid too much complexity? What tools are helpful and how can we improve usability? This paper will not provide definite answers to these questions, but it gives an insight into the difficulties, challenges and risks of using interfaces to investigate historical newspapers. More importantly, it provides ideas and recommendations for the improvement of user interfaces and digital tools.

**INTRODUCTION**

Interfaces for digitised, historical newspapers offer great opportunities for many different user groups, such as academic researchers, lay historians, students, teachers, etc. Interfaces steer what users can learn from digitised newspapers and they influence workflows by offering functions and tools [cf. Jarlbrink and Snickars 2017]. On the other hand, users are often not aware of biases in the search results which are caused by the processing and datafication of newspapers. The main aim of this paper is to discuss the opportunities and pitfalls of research based on digital newspaper interfaces. We argue that integrating tools for analysis and visualisation of data and metadata would not only address the needs of various user groups, but that libraries could also profit from expanding their digital interfaces.

In order to be able to make such assertions researchers, of a DH team[1] from Austria, Finland and France in the EU-funded research project 'NewsEye: A Digital Investigator for Historical Newspapers'[2] experimented extensively with existing newspaper search interfaces of the three national libraries using specific humanities case studies. The interfaces used were ANNO (ONB

---

[1] The following researchers contributed to this essay: University of Innsbruck: Eva Pfanzelter, Sarah Oberbichler, Barbara Klaus and Stefan Hechl; University of Helsinki: Jani Marjanen and Mikko Tolonen; University Paul-Valéry Montpellier: Marie-Eve Thérenty, Pierre-Carl Langlais and Nejma Omari; University of Vienna: Martin Gasteiner.

[2] This work has been supported by the European Union's Horizon 2020 research and innovation programme under grant 770299 (NewsEye).

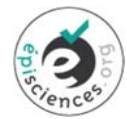

– Austrian National Library), Digi (National Library of Finland) and Korp (The Language Bank of Finland), as well as Gallica and Retronews (BNF – National Library of France).

In addition, the involved DH teams set out to use various tools that are publicly available online to analyse the newspaper corpora outside the interface environments. Since the corpora are way too large to export, they created subcorpora of the data they used for their research – a task that turned out to be one major and tricky first step in newspaper research. Using these subcorpora, they then explored different ways of performing frequency analyses, topic modelling and context analysis.

[Ehrmann, Bunout and Düring, 2019] published a survey on 24 interfaces for digitised historical newspapers with the aim of mapping the current state of the art and identifying recent trends with regard to content presentation, enrichment and user interaction. In this survey, they confirmed the continuous existence of a gap between growing user expectations and current interface capacities. While most interfaces allow users to search, filter and view newspapers, there is still a lack of content management and enrichment by NLP methods. This conclusion also applies to the three interfaces examined in this paper. However, on top of these results, the NewsEye DH team tried to find out how far they could advance their own research with the existing interfaces, and where problems, distortions and limitations would arise. In contrast to the study by Ehrman et al., which tested interfaces on the basis of a list of criteria to be able to compare them, this paper uses three case studies by historians and linguists in order to show specific research examples. This should help to illustrate problems and to better understand the approach of humanities scholars. Therefore, our aim is not to compare these interfaces or argue for a common gateway, but to show the difficulties that come up when working with them for different user groups.

As an overall result, it can be stated that the NewsEye DH team was able to draw some important conclusions: Without the possibility to create subcorpora, humanities researchers often cannot align their analyses with their specific research questions. Keywords are still key, and frequencies should be state of the art for contextualisation and content analysis not only for academic users. Also, topic modelling and word embeddings hold many promises for future development.

So far, OCR quality, faulty images and metadata as well as missing functions for analysis and visualisation complicate and restrict the use of the newspaper interfaces. On the other hand, existing tools that are not integrated into the interfaces often lack transparency, user-friendliness and the possibility to process large corpora of messy data. In this sense, although some interesting hypotheses could be generated and some results could be used in support of existing studies, there is ample opportunity for improvement, if newspaper interfaces are to be used as reliable gateways to in-depth analysis of these extensive archives of cultural heritage.

**I SURVEY ON DIGITAL NEWSPAPER USAGE**

As a part of the wider effort to examine the interest in, and use of, digitised newspaper collections across Europe, as well as to find out how the general public interacts with them, the NewsEye DH team together with the national libraries of Austria, Finland and France in 2018 and 2019 conducted user surveys on the different online interfaces offered by the libraries. All in all, the surveys showed an acceptable level of happiness with the basic functionalities of the newspaper interfaces. While it appears that the user interfaces offer interesting tools, not everyone finds them easy to use. The feedback regarding this issue can be summarised as follows:
- Advanced search features sometimes require additional explanations and are hard to understand, especially for inexperienced users.

- It also became clear that the advanced keyword search preferred by experienced users is not accepted outside academia as well as expected .
- Respondents made extensive use of filter functions, and they are also very keen on the download options.
- Interestingly enough, many respondents are aware of limitations of the interfaces and have many suggestions for additional tools that could be implemented.
- While some users do not seem to know some of the vocabulary used in the survey (OCR, metadata, full text search, etc.), they were bothered by (OCR) errors in their findings.
- Although respondents would like many features to be implemented in the search pages, they are not willing to pay for such a service.

These findings based on the wishes of users were very much in line with the experiences of the NewsEye team, which in May and June 2019 carried out extensive testing of the aforementioned interfaces. The overall results of these sessions are condensed in the following paragraphs.

**II INTERFACES, METHODS, AND TOOLS**

In order to be able to compare results, the NewsEye DH team first focused on thoroughly using the existing newspaper interfaces in order to further their insight into some of the historical case studies they are conducting and, secondly, used tools and methods openly available on the internet. The starting point, however, were the newspaper interfaces.

The interfaces the humanities researchers were working with are part of the Austrian, Finnish and French national libraries:

|  | **ANNO** | **Digi and Korp** | **Gallica and Retronews** |
|---|---|---|---|
| **Provided by:** | Austrian National Library | National Library of Finland and The Language Bank of Finland | French National Library |
| **Offers:** | Free and open access to 21 million Austrian newspaper and magazine pages, which were published between 1568 and 1948. | Both, Digi and Korp provide data dumps of the newspaper material that are openly available up to the year 1919. Digi contains over 16.6 million pages of digitised material. Korp provides a slightly smaller version with morphological information added to the text. | While Gallica provides access to all digital assets of the national library (including 15.937 newspaper issues), Retronews is exclusively dedicated to the historical press and offers 9 million digitised pages dating from between 1631 and 1945. |
| **Links:** | http://anno.onb.ac.at/ | https://digi.kansalliskirjasto.fi/etusivu <br> http://korp.csc.fi | https://gallica.bnf.fr/ <br> https://www.retronews.fr/ |

Table 1. The newspaper interfaces of the Austrian, Finnish and French national libraries.

**2.1 Case studies on migration, nationalism and gender**

The DH researchers then set out to work on three different case studies, which were used as a basis for the testing of the tools and the interfaces.

|                              | **Migration**                                                                                                                                                                                  | **Nationalism**                                                                                                                                                                                                    | **Gender**                                                                                                                                                                          |
|------------------------------|------------------------------------------------------------------------------------------------------------------------------------------------------------------------------------------------|--------------------------------------------------------------------------------------------------------------------------------------------------------------------------------------------------------------------|-------------------------------------------------------------------------------------------------------------------------------------------------------------------------------------|
| **Short description:**       | The chosen subtopic on return migration is part of a case study on migration and attempts to reveal return migration processes to Europe between 1850 and 1950                                 | The case study on nationalism focuses on the changing language of nationhood in the 19th and early 20th century.                                                                                                   | The case study on gender aims to bring to light the evolution of the representation of women in the news from 1850 to 1950.                                                         |
| **Research questions:**      | How and in what context were Austrian daily newspapers reporting on returnees and how did the reporting change over time?                                                                      | How did discourses on nationhood expand in the 19th century? How and why did the notion of nationalism become a central figure of thought in the early 20th century?                                               | What place is given to women in the newspapers? What role do newspapers play in the struggles of women?                                                                             |
| **Digital tools used and why:** | The case study uses topic models and word embeddings in order to find, select and organise articles for further qualitative analysis (discourse and argumentation analysis)                 | The case study combines reading of sources with frequency analysis and word embedding models to cluster key terms.                                                                                                 | Digital tools (notably topic modelling and frequency tools) as well as named entities are very valuable in this case study for revealing the trends and key figures.                |

Table 2. Case studies on migration, nationalism and gender.

## 2.2 Creating a subcorpus

Considering the amount of hits usually received when performing keyword searches on the newspaper interfaces, it quickly became clear that it is imperative for researchers to be able to create subcorpora of the available datasets. Limiting the number of results, however, is imperative when humanities researchers work qualitatively – a research approach that can never be ignored in these disciplines. None of the tested interfaces offer tools to select, structure, or build datasets, and the DH researchers were forced to create their own collections manually, an extremely time-consuming task. Methods to support the automatic creation of collections, e.g. using text mining methods, are still experimental and can at this point not be used on newspaper interfaces. [Nanni, Ponzetto and Dietz, 2017], for example, used named entities and word embeddings to automatically build entity-centric event collections in order to address the needs of humanities and social science scholars who work qualitatively on specific topics. The newly developed interface for historical newspapers, 'Media Monitoring of the Past' [Impresso, 2019], is taking a first step towards supported collection building. The approach there is to connect topic modelling with search queries in order to tailor search results to a specific topic.

2.2.1 Subcorpus on return migration (1864–1944)

When researching processes in society rather than strictly defined topics, humanities scholars often work with specifically pre-selected subcorpora of the sources they are investigating.

Questions of remigration – and migration in general – are among those phenomena that come hand in hand with several challenges: First, there are no specific keywords that would help to identify the articles needed for a study of past migratory movements, and/or the words addressing these issues have changed over time. Also, even though distance searches allow users to search for terms appearing together in a specific surrounding, there is no guarantee that words appear in the desired contexts. Second, return migration was (and is) not widely talked about in newspapers. Therefore, methods such as word frequencies or topic modelling can hardly be used when looking for a suitable subcorpus, because due to the relatively small amount of hits, the desired sources would remain hidden in the mass of the digitised newspaper corpora. Third, without knowledge of the historical context, the research area, and the factors of changing language use over longer time periods, it is not possible to decide whether a subcorpus is suitable for the research in question.

For the case study on return migration, a subcorpus from the ANNO dataset was created, focusing on questions about types of returnees, nationalities, countries, and other topics related to remigration. As a first steps, the keywords in table 3 were used to find articles in ANNO about remigration issues:

| keywords | |
|---|---|
| *'heimat zurückkehren'*~20 ('returning home') | *'heimat rückkehr'*~20 ('return home') |
| *'Heimkehrenden'* ('returnees') | *'Heimgekehrten'* ('returnees') |
| *'emigranten rückkehr'*~20 ('emigrants return') | *'Heimkehrer'* ('returnees') |
| *'Rückwanderer'* ('returnees') | *'Rückkehrer'* ('returnees') |

Table 3. Keywords used for the subcorpus on return migration in ANNO.

The subcorpus finally contained 472 articles on remigration issues from 1864 to 1944, with 109.031 total words and 22.193 unique word forms, divided into 77 documents (each document covering one year). The subcorpus is based on articles from two newspapers available in ANNO: *Neue Freie Presse* (1864–1944) and *Illustrierte Kronen Zeitung* (1908–1944). In the selection process, the relevance of the articles for the topic and the OCR quality were taken into account. Some errors were corrected manually, and many line breaks were removed. Overall, however, the original text quality was maintained. The articles were copied from the interface and saved as TXT, DOC, and CSV files.

Due to missing metadata, a graph of absolute frequencies of the subcorpus on return migration had to be created manually (figure 1). It has to be noted that while the *Neue Freie Presse* was published throughout the period studied, the *Illustrierte Kronen Zeitung* was printed between 1908 and 1944 only. Despite this, due to OCR problems, hardly any articles could be found in the *Neue Freie Presse* after 1930. In the *Illustrierte Kronen Zeitung*, text recognition was consistently good and the findings reasonable.

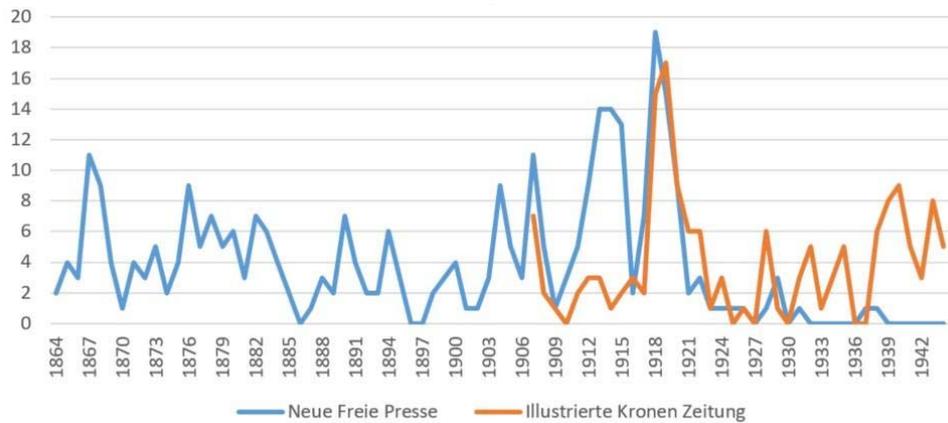

Figure 1. Absolute frequencies of the subcorpus on return migration.

The selected articles deal with different kinds of return migration. They were chosen to cover different types of reporting and content in order to allow different research questions to be answered. However, the following challenges arose during the selection process:
- Keywords: Especially the keyword search *'heimat zurückkehren'*~20 ('return home') and *'heimat rückkehr '*~20 ('returning home') led to many results that were not related to return migration. Many newspapers and articles had to be excluded manually.
- Time: Creating this small corpus containing nearly 500 articles took almost 50 hours.
- Workspace: Since there is no specific place on the ANNO interface that allows for a collection and/or further analysis of found articles, all the material had to be downloaded one by one and processed using other software. Metadata and context are lost in the course of this process.
- Metadata: While information about the newspapers itself (title, publication year, etc.) can be found in ANNO, details about the content (number of words, pages, languages etc. per year/month/day, document length) are not available.

2.2.2 Subcorpus on 'women's suffrage' (1890–1940)

The subcorpus on woman suffrage contains 628 articles from *Le Matin* and *Le Petit Parisien* published from 1890 to 1940 and available in the BNF newspaper dataset. The corpus includes the articles containing the French expression *'vote des femmes'* ('women's suffrage') in these two daily newspapers (figure 2).

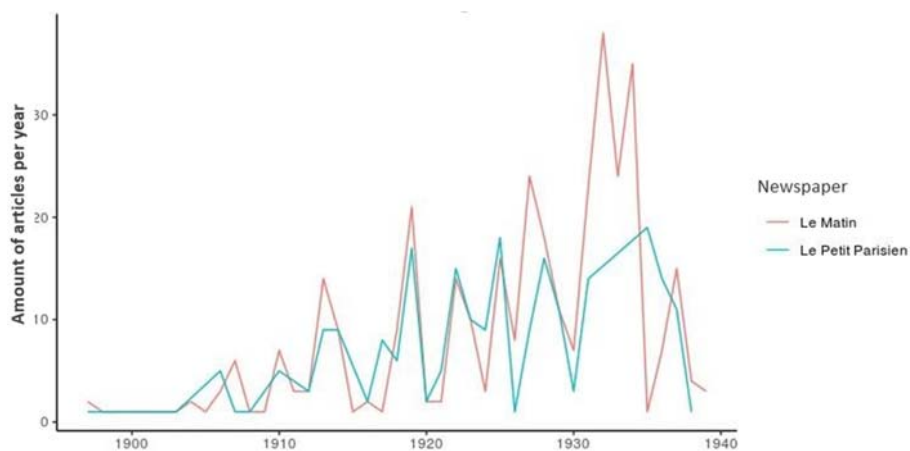

Figure 2: The articles of the French women suffrage corpus per year in *Le Matin* and *Le Petit Parisien.*

In this case, it was possible to extract articles thanks to the Optical Layout Segmentation done by the Europeana Newspaper projects in 2011–2014: Paragraphs and text blocks were grouped into articles. This digital reconstruction of articles is not perfect and is especially faulty for advertisements, which tend to be printed in messy sections, regardless of the actual segmentation between ads. Nevertheless, standard articles with a title are recognised correctly enough in order to be suitable for research purposes.

The complete digital archives of the two newspapers were already available within the programming interface of the *Numapresse* Project, hosted on the French national digital infrastructure for the humanities and social sciences, *Huma-Num*. The original dataset includes numerous supplementary pieces of metadata at the word level, for instance, the coordinates, the size or the font style for each token. This contextual approach to text mining has been described by [Langlais, 2019]. Here, only the aggregated raw text for each article published in *Le Matin* and *Le Parisien* from 1890 to 1940 was used and the 628 publications with the string *'vote des femmes'* retained. This selection is only a small portion of available material and shows a tiny sample of the journalistic debate on women's suffrage. To get a wider perspective, it would have been necessary to use many alternative expressions (e.g. *'suffrage féminin'*) or neighbouring searches (such as targeting all the articles where the word *'femme'* is close to the word *'politique'*).

The quality of text recognition can affect the results as well. OCR accuracy is usually between 80–90 % for this period, but can go as high as 95–99 % during the 1930s, probably thanks to an enhanced quality of the original news archive. This discrepancy of text recognition across the period can potentially skew the results: We may have more articles from the 1930s simply because the query matches more results.

As the original purpose of this part of the project is to experiment with several tools and not to produce a general inquiry from a social science perspective, a smaller corpus seemed appropriate.

2.2.3 Subcorpus on nationalism

The subtopic on nationalism focuses on the long-term changes in the vocabulary relating to the nation in Finnish newspapers, with a starting point in 1771 and ending in the 1920s. Simple free-text searches yield up to two hundred thousand hits for the words *'nation'* in Swedish and *'kansakunta'* in Finnish. Hand curating a subcorpus of articles based on the search results would be possible only if the selection was to be narrowed down remarkably to a particular aspect of writing about the nation, as in some of the previous subtopics. The path chosen for this subtopic was rather to choose a selection of empirical cases that were studied qualitatively, but then doing the bulk of the analysis by using descriptive statistics with all the instances containing words belonging to the vocabulary of the nation. Some of the results are described below under the heading 'Frequency analysis'.

However, for some of the analyses produced for the subtopic, subcorpus creation was in effect done by using search terms and downloading the Key Words in Context (KWIC). Contexts were created either according to a set window (five or more words before and after the selected term) or according to a flexible context (downloading whole paragraphs or sentences in which a particular word appeared in which cases the length of the window varies). These subcorpora were used in particular when producing word-vector-space models for analysing in which contexts the vocabulary of the nation was used in Finnish newspapers [Hengchen et al., 2019].

They were also used to produce frequency lists of the lexical context of 'nation', 'national' and 'nationalism'.

The method of focusing on a set window or flexible context is not as accurate as using identified articles as context for analysis. As the Finnish newspaper dataset does not include reliable article segmentation, this, however, remains our best method of doing contextual analysis without hand-curating the subcorpus. Using a set window has the benefit of producing symmetrical contexts for each keyword search, but this method is ignorant of sentence structure or stylistic features in the text. If the window is large, say one hundred words before and after, it is also more likely that the chosen context includes text from articles that precede or succeed the target article. Focusing on a sentence or a paragraph has the benefit of rarely transgressing article boundaries, but this method sometimes gets the context wrong due to OCR errors regarding punctuation. Furthermore, focusing on the sentence or paragraph as context gives more prominence to articles with long sentences or paragraphs. For these reasons, the set window worked with is fairly small.

**2.3 Frequency analysis**

For historians working with large quantities of digitised texts from longer periods of time, the possibility to not only note the introduction of new vocabulary, but also to be able to quantify the frequency of words or other features of language has become a way of illustrating changes in how people conceptualised the world. Newspapers form an especially interesting data set for this, as they are, relatively speaking, a rather stable corpus that deals with a large selection of topics. The first step when working with an adequate corpus using digital tools is therefore usually a simple frequency analysis. Next to co-occurrences or topic modelling tools, this can be a first entry point for more in-depth investigations of larger and complex datasets, since it can give researchers a quick overview of word frequencies (or at least frequencies of strings) and trends in the selected dataset.

It is common practice in humanities scholarship to trace the first occurrence of a word relating to the topic of a study. As [Dufuoix, 2019: 15] puts it, this can be called 'the religion of the first occurrence'. Digitised sources are obviously helpful in identifying new candidates for first uses of particular terms through keyword searches, but more importantly, they make it possible to shift focus from early uses to the analysis of frequency and answering questions of when it became more common in general to use a particular word or when only some people chose to use a particular terminology.

Digitised newspapers are a very good dataset for diachronic analysis of word frequency for many research questions, as it is relatively easy to select subcorpora based on particular newspapers (perhaps with different political leanings), particular places of publication, or the publication frequency of newspapers. To compare changes in word frequency over time, frequencies need to be normalised to account for different sized corpora for each time slice (usually a year), so that we get a relative frequency of words used per one million (for instance) words in the corpus of each year.

For corpus linguistics, the analysis of word frequencies is much about getting to the truth of language use, and hence having a balanced corpus is imperative. For historical purposes, more uncertainty is acceptable and also unavoidable because of the nature of historical research questions. The historians' interpretation about what changes in word frequencies are about must also try to account for changes or biases in the data. In general, changes in word frequencies are about (but not limited to):
- a word becoming very topical in a given moment
- a word entering new domains in language

- a word being included in larger linguistic entities, such as idioms, which may have a meaning of their own
- a word is associated with new related meanings (polysemy)
- something data-specific that skews frequency. In the case of newspapers, this might, for instance, be repeated advertisements that are surely part of the historical dataset, but would be cleaned out if the data set was seen as balanced in a corpus-linguistic sense.

Of these examples, the issue of polysemy is perhaps the most complicated one, as semantic change can happen in many different ways and be about rather different things. For instance, semantic change might be about new things (referents) that need to be described, but might also be about changes in the valuation of a word, or, indeed, something else. If we take into account different types of meaning ranging from conceptual meaning to social meaning and affective meaning [Leech, 1974] and also different types of reasons for change in them [Ullman, 1962], we end up with a rather complex model for lexical semantic change. Changes in word frequency are not helpful in assessing them, but they are useful in trying to identify when such changes took place.

Taking the importance of frequency analysis into account, the DH researchers working on their case studies used the newspaper interfaces in Austria and Finland in order to find out where the chances and limitations of the method are.

2.3.1 Frequency analysis on nationalism using Digi and Korp

The subtopic on nationalism relies on digitised newspapers as a massive text dataset that can be used to trace long-term changes in the vocabulary of 'nation', 'national', 'nationalism' and related terms. As such, it provides an example of the possibilities and limits of using frequency analysis in the digitised Finnish newspapers to study complex historical phenomena.

One first step is to understand that words like 'nation' and 'nationalism' should not be studied alone, but in the context of related terms, but at the same time it is clear that individual words do have their own trajectories and that choosing to talk about 'nationalism' and not just 'nation' has most probably been a conscious choice by a historical actor. Typically, the first uses of the words 'nation' and 'nationalism' have been mapped thoroughly, for instance by [Kemiläinen, 1964]. One of the most famous and early uses of the word 'nationalism' is by [Johann Gottfried Herder, 1774] in *Auch Eine Philosophie der Geschichte zur Bildung der Menschheit* (in English: 'This Too a Philosophy of History for the Formation of Mankind'). Herder writes about the animosity between nations and how Herder's philosophical/political opponents label this as 'prejudice! mob-thinking! limited nationalism!' As Herder is one of the main figures in the history of national thought, this use of 'nationalism' has naturally been noted, but from the point of frequency of the use of the word 'nationalism' e.g. in Swedish-language newspapers, the issues come across differently. Early uses, such as the one by Herder, appear as being occasional uses, whereas 'nationalism' becomes more frequently used only at the end of the nineteenth century and beginning of the twentieth century (figure 3).

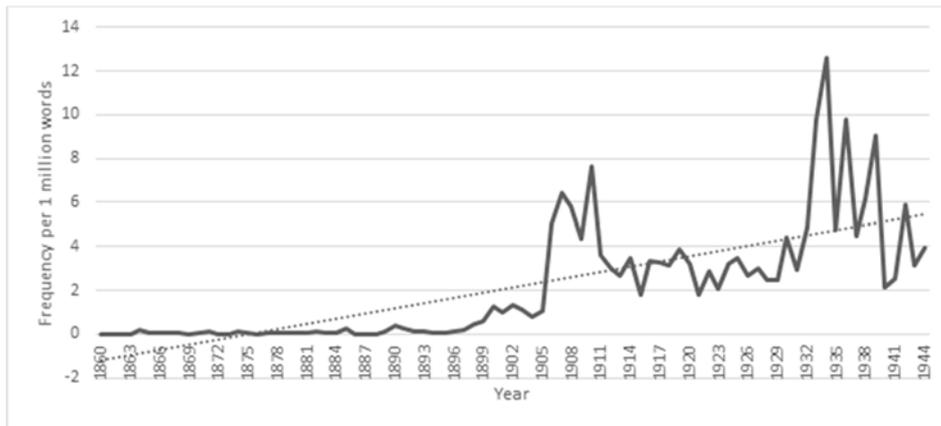

Figure 3. Relative frequency (hits per one million words) of the lemma 'nationalism' in Swedish- language newspapers in Finland. False positives resulting from bad OCR of the word 'rationalism' as 'nationalism' are omitted.

A reading of some of the uses of 'nationalism' in the late nineteenth century shows that the word was used in a negatively laden way in heated debates about Swedish, Finnish and Russian nationalism in Finland. Also, the rise in frequency of the term 'nationalism' at this time was not only a Finnish phenomenon, but can be seen in other similar datasets, such as the manually created absolute frequency in figure 4 from ANNO. The growth in frequency for 'nationalism' was clearly about the word becoming topical in certain debates.

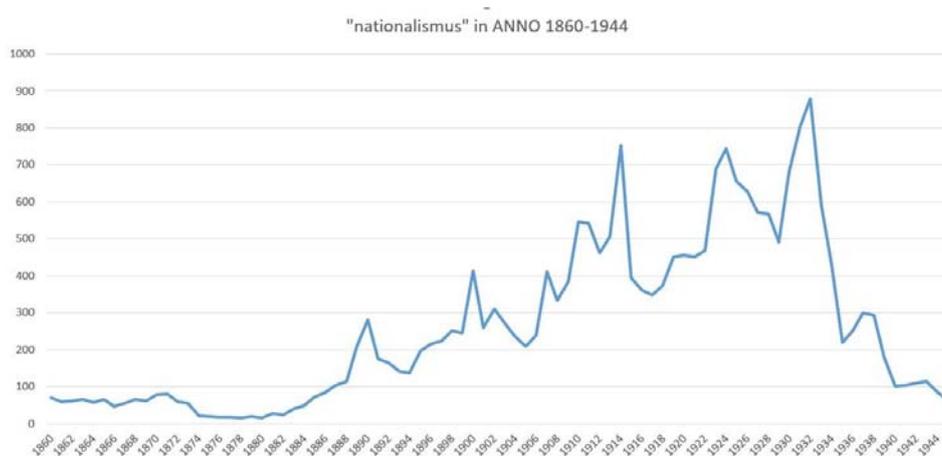

Figure 4. Absolute frequency of the keyword *'nationalismus'* in German language newspapers in Austria. Early hits include many false positives due to OCR errors in reading the word *'rationalismus'* like in the Finnish case. Bad OCR is also corrupting the results from 1930s onward (see below).

If we turn to the word 'national', the story is also a story of growth during the nineteenth century, but still slightly different (figure 5).

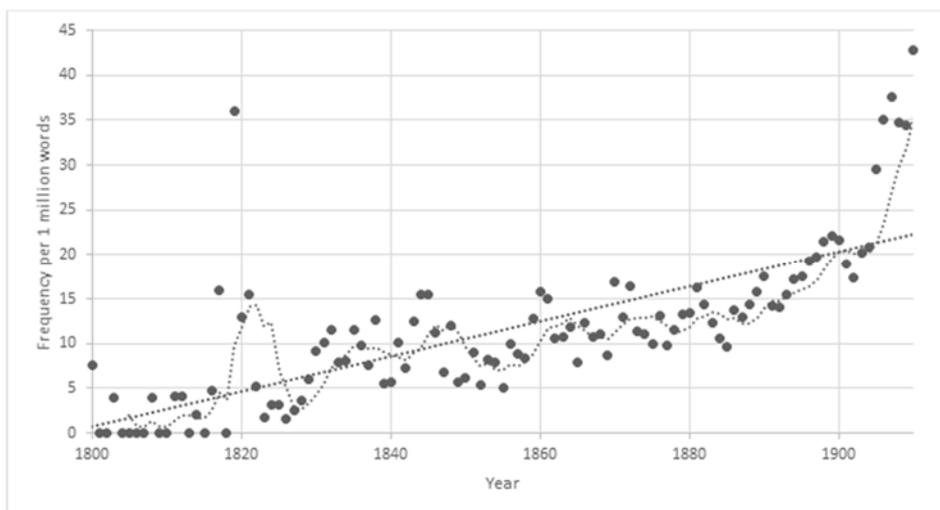

Figure 5. Relative frequency of the lemma '*nationell*' in Swedish-language newspapers in Finland, 1800–1910.

The fact that the relative frequency of 'national' and 'nationalism' do not quite correspond suggests that they address slightly different things. Hence, also the trends in their use differ. Looking at not only the frequency of 'national', but also looking at the context of the term is revealing in this sense. Since 'national' is most often followed by a noun that tells us which things could be conceptualised as being national, we can also count how many different nouns co-occur with 'national' each year in the newspaper corpus. This is called the productivity of 'national' (figure 6).

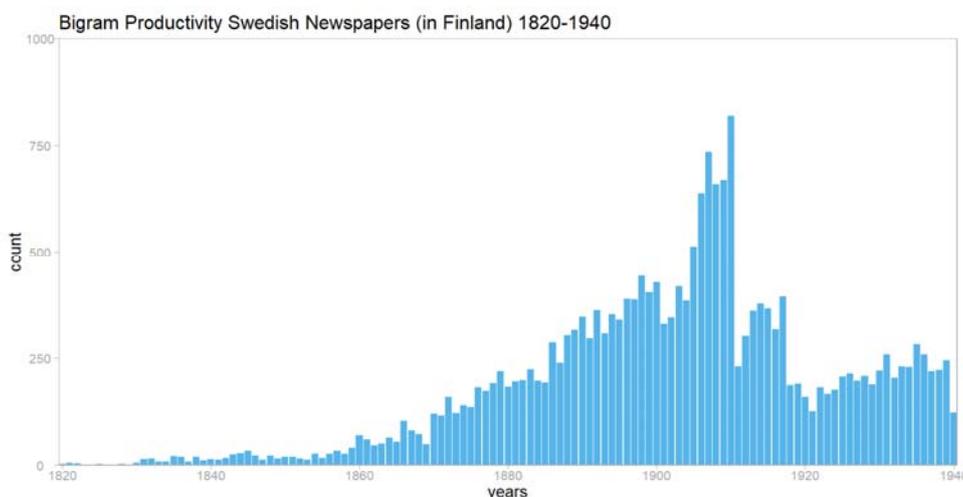

Figure 6. Productivity in the bigrams starting with the word '*nationell*' ('national') in Swedish-language newspapers and periodicals in Finland, 1800–1945. After 1910, the size of the annual data sets is smaller. (The graph was produced by Ruben Ros.)

The increase in productivity indicates that over the course of the nineteenth century, the word 'national' was used in more and more domains of language. By the end of the century, the nouns co-occurring with 'national' related quite different spheres of life from 'national economy' and 'national assembly' to things like 'national anthem' or 'national language'. The proliferation of nouns co-occurring suggests that also the general growth in relative frequency is more about the national perspective becoming nearly all-encompassing. Only in the years 1820 and 1821 do we see a clear peak in relative frequency that can be explained by a factor in the data. In those years, the publications *Mnemosyne* and *Åbo Underrättelser* radicalised the language of 'national', which is also clearly visible in the data.

Frequency analysis for the purpose of historical interpretation still relies on a heavy interpretative component. Historians are often interested in changes in language, but also relating those changes to societal change, which always requires a discussion about how changes in language use relate to other changes in society. In the cases of the words 'national' and 'nationalism', we see that while the words share a root, the concrete uses of the terms differ. As pointed out by [Kurunmäki and Marjanen, 2018], this is much due to the rhetorical function that is very often accompanied with the use of different 'isms'. Studying the frequency of the word 'nationalism' is thus not a good indicator for studying how people understood the nation or even less how nation-building as a process advanced in Finland (or elsewhere), whereas looking at the frequencies of 'national' tells us about the process in which the national perspective became dominant and therefore also relates more closely to nation-building as a process.

For historians to better understand to which extent relative frequency can be used to draw conclusions relating to historical processes and get beyond some of the naive interpretations that have emerged especially after the Google Ngram Viewer was made public [Pechenick, Danforth, and Dodds, 2015], they need better tools for using frequency measures for explorative work. As producing frequency graphs is (computationally) not a demanding task, frequency analysis should not be left to the projects that can download full data of digitised data sets and process them, but should ideally also be available as integrated tools in graphical user interfaces – as is the case, for instance, for digitised Dutch newspapers at [KB Lab].

2.3.2 Frequency analysis on return migration using ANNO

Since the scholarly literature dealing with the issue of return migration to Europe is limited, little is known about remigratory processes between 1850 and 1950. A frequency graph could therefore reveal when remigration became an important issue in the newspaper coverage, and, as a result, an issue for the societies people returned to.

ANNO does not yet provide any tools to create frequency analyses automatically. Nevertheless, with the help of the information provided by ANNO (absolute number of pages with search terms in a certain period of time) it is possible to create frequency graphs by manually copying the numbers to a spreadsheet outside the platform. The absolute numbers of hits per year are displayed in the filter section of ANNO and can be extracted manually (figure 7).

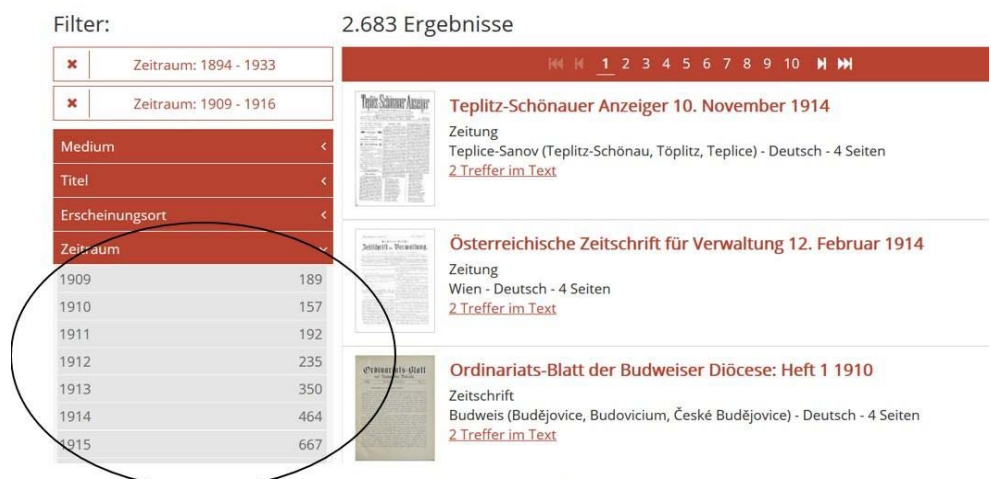

Figure 7. Hits of the search term *'Heimat zurückkehren'*~20 in ANNO.

After entering the obtained hits into an Excel file, an absolute frequency graph for the search terms *'Heimat'* in combination with variations of *'rückkehr /zurückkehren/heimkehren/zurückgekehrt'* ('home' combined with 'returning/return/return home/returned') could be created (figure 8). Since the distance search function does not allow terms to be marked with an asterisk (*), many term combinations were required (e.g. it is not possible to search for *'heimat *rückkehr*'*~20/'home *return*').

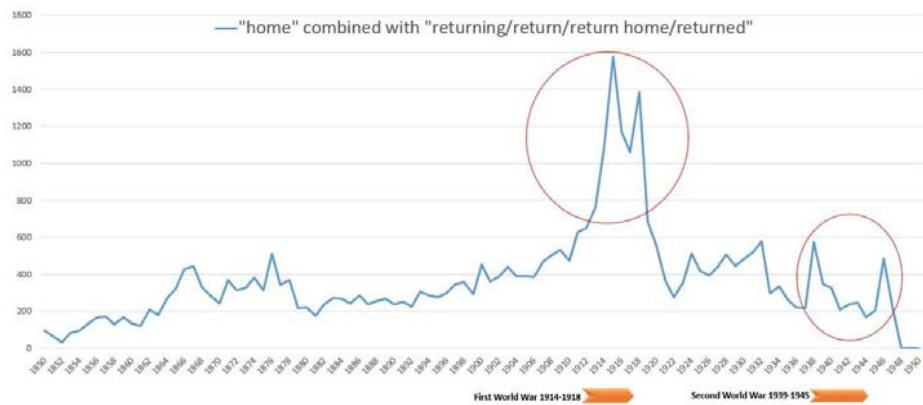

Figure 8. Frequency graph of 'home' combined with 'returning/return/return home/returned' (1850–1950).

Taking historical events into account, a closer look at the graph immediately revealed a discrepancy: The number of hits rises sharply – somewhat expectedly – during and at the end of the First World War, but very surprisingly not during the Second World War. This discrepancy also became apparent with many other search terms. The question of why the results of the frequency analyses were distorted led to a consideration and subsequent examination of the following questions:
- Were important search terms overlooked?
- Is there a discrepancy between the number of newspapers published during the First World War and the Second World War?
- Is there a discrepancy between the number of digitised newspapers available from the First World War and the Second World War?
- Is there a change in the quality of the digitised newspapers or the OCR for certain newspapers?

In order to facilitate the interpretation of the results, the same keyword search ('home' in combination with 'returning/return/return home/returned') was carried out by selecting only one newspaper. This made it easier to estimate the quality of the OCR, the number of issues published, and the volume of the dataset per newspaper. The newspaper *Illustrierte Kronen Zeitung* (1905–1944) seemed to be suitable for this, as the quantity and the OCR quality remained more or less the same throughout the entire time period. As it turned out, however, there is no OCR text available for the years 1936 and 1937. This limited search delivered different results. In contrast to figure 8, figure 9 shows a clear increase in the usage of the search terms at least for the beginning of the Second World War:

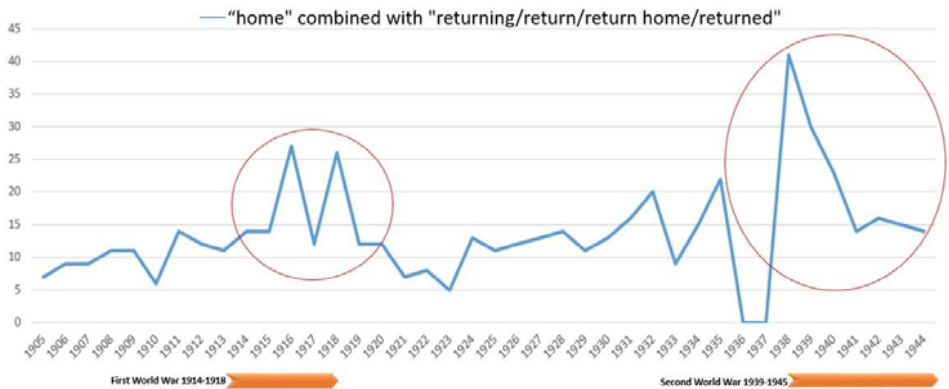
Figure 9. Frequency graph of the search term 'home' combined with 'returning/return/return home/returned' in the newspaper *Illustrierte Kronen Zeitung*.

Therefore, figure 9 reveals that missing keywords are not the main problem here. Instead, two issues were identified:
- Absolute frequencies can show an unrealistic picture, as they only count the number of times a word or a page containing the word appears in a corpus without putting the results in relation, e.g. with the total number of words or pages for a certain time period or a certain newspaper. However, relative frequencies cannot be generated with ANNO data as it exists now, because required metadata is not available, i.e. word counts or number of digitised pages/newspapers per year.
- Very time-consuming browsing and comparison of the two selected newspapers and their OCR'd versions revealed that the quality of the OCR between 1938 and 1945 varies extremely from newspaper to newspaper. Figure 10, for example, shows an article on return migration in the newspaper *Neue Freie Presse* from 1938. On the left, readers can see the scan of the original text and easily read it. On the right, the OCR'd letters do not resemble German words at all. Apart from the headline, almost no word is readable and many of the characters shown are not part of a normal German text.

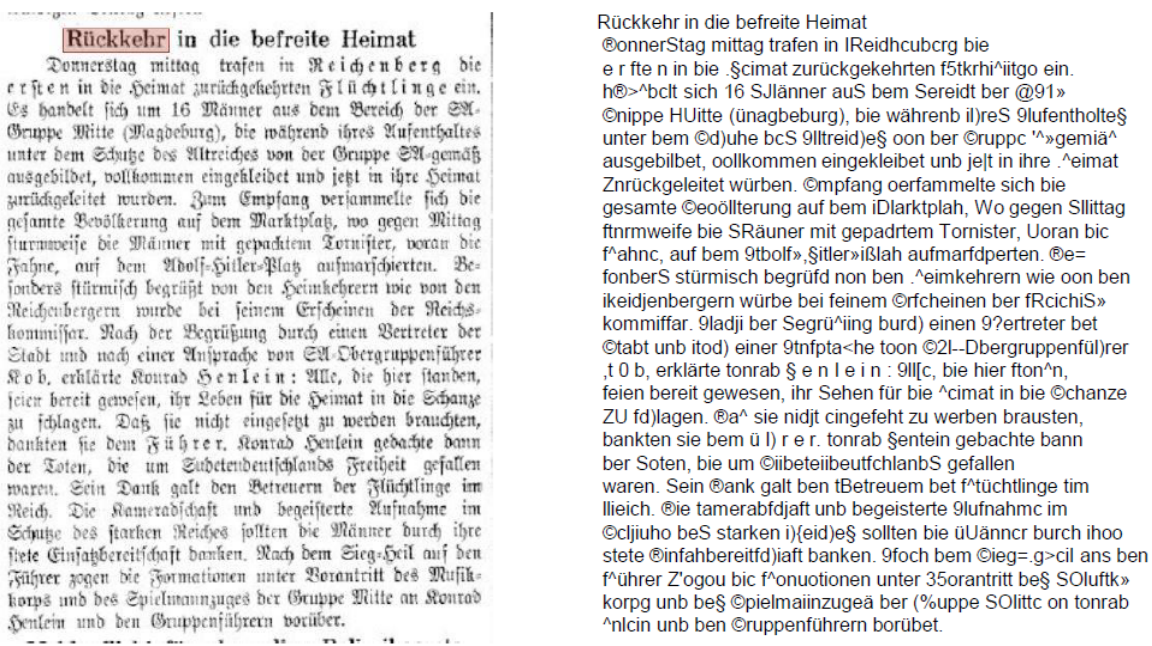
Figure 10. *Neue Freie Presse*, October 15th, 1938 (PDF, left / TXT, right).

To summarise, it can be said that the OCR quality in ANNO varies significantly not only over time, but also between different newspapers. Also, it became clear that the lack of metadata and the lack of adequate tools prevents the preparation of conclusive frequency analyses, which in turn could point users - and libraries that host the interfaces - to some of the flaws mentioned above. Still, the manual creation of frequencies is time-consuming and cannot seriously be considered for ordinary research efforts. Last but not least, the irregularities found in the search results lead to a more or less devastating conclusion, namely that the automatically created search results can be invalid, biased or even false.

**2.4 Analysing created subcorpora using existing topic modelling tools**

For historians, computer science and digitisation have held many promises. From the early 1970s until today, however, as [Blevins, 2016] argued, the 'sunrise of methodology was still hovering just over the horizon' for digital history. Digital methods, despite all the advances in recent years, especially concerning quantitative approaches, seem to be an eternal promise. One of those promises certainly is the potential possibility to instruct machines to automatically extract meaningful topics from huge corpora of data. It has sparked the discipline's interest in the computer scientists' topic modelling approaches ever since the introduction of the concept in the late 1990s, as described by [Papadimitirou et al., 1998].
Topic modelling tools can basically be understood as algorithms that extract meaningful topics from text. According to a tweet by [Garfinkel, 2012], a topic can be defined as 'a recurring pattern of co-occurring words'. Or, in the words of [Graham, Weingart, and Milligan, 2012], a so-called topic 'consist of a cluster of words that frequently occur together'. In a good topic model, the results – the topics or rather the words in the topic – make sense to the reader. [Brett, 2013], for instance, uses the topics 'tobacco, farm, crops' and 'navy, ship, captain' as examples. Newspapers have been a popular subject for topic modelling, because they provide a way to study the change of words over time from a daily source, as [Brett, 2013] argues. Current and historical newspapers have been investigated with topic modelling, see for instance the work by [Nelson] on *The Dispatch*. At the same time, not many topic modelling tools provide humanities or social science researchers with the techniques they need in order to meaningfully interpret the recurring pattern of co-occurring words. Even though there are attempts to add structure to large topic models and to make them approachable and useful to an end user through interactive visualisations [cf. Smith, Hawes and Myers, 2014 / Cai, Sun and Sha, 2018], the most useful function – a display of where the topics occur in the corpus – is mostly missing.
Since there are no topic modelling features available on the newspaper interfaces investigated here, the NewsEye DH teams used several other available tools that allow users to experiment with topic models, namely STM, TidySupervise and Overview:

| **STM** | **TidySupervise** | **Overview** |
|---|---|---|

| | | | |
|---|---|---|---|
| **Description** | STM is centred on non-supervised text classification and does not focus on corpus preparation. STM is fairly new and documentation is slightly lacking: There is little explanation on the package page and for the current case, the well-commented use by [Silge] was mostly relied on. | TidySupervise is a new tool recently developed by the *Numapresse* project. It originally started as a set of customised functions to deal with the most ambitious project of *Numapresse*: the classification of newspaper per news genre for a very long period of time (1895–1940 for both *Le Matin* and *Le Petit Parisien*). | Overview promises to be able to automatically analyse thousands of documents. Apart from text to topic clusters, it includes full text search (including fuzzy search), visualisations and entity detection. |
| **Form** | R package | R package | document mining application |
| **Usability** | Needs basic programming knowledge | Needs basic programming knowledge | No programming knowledge required |
| **Links:** | https://cran.r-project.org/web/packages/stm/index.html | https://rdrr.io/github/Numapresse/TidySupervise/ | https://www.overviewdocs.com/ |

Table 4. Topic Modelling Tools STM, TidySupervise and Overview.

2.4.1 Creating topics on return migration using STM – an R package for the structural topic model

STM was used with the corpus on return migration (472 articles). The corpus was prepared using naive lemmatisation with lookup tables: This is not a significant issue for text classification based on thousands of individual tokens. Stop words were not removed manually. The following eight topics were generated for the whole corpus (figure 11):

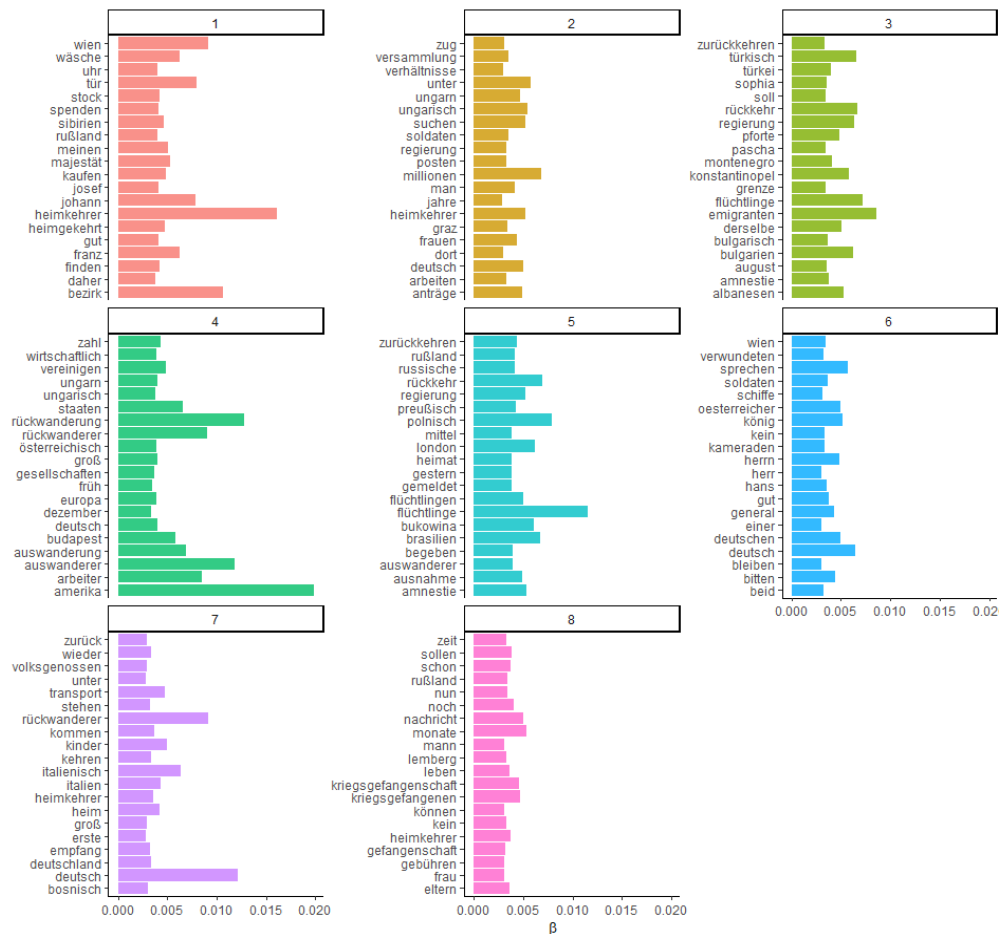

Figure 11. Topic model with STM from the subcorpus on return migration[3].

At a first glance and without being familiar with the content of the corpus, it is difficult to tell what the topics really stand for. In order to provide more accurate interpretations, a better insight into the corpus is needed. After reading, the following conclusions can be reached:
- Only topic 4 seems to be very clear: the return of people from America.
- Topic 1 seems to map the discourse of laundry donations for returnees.
- Topic 8 seems to reflect the discourse on prisoners of war.
- In topic 3, the question arises of how the words '*türkisch/Türkei*' ('Turkish, Turkey'), '*Sophia*' ('Sofia'), '*Montenegro*', '*bulgarisch/Bulgarien*' ('Bulgarian, Bulgaria'), '*Albanesen*' ('Albanians'), '*Konstantinopel*' ('Constantinople'), '*Flüchtlinge*'

---

[3] Translation of topics (the translation of German terms may not always be accurate, as the translation would require more context. STM does not allow to see words in their original context).
**1**: vienna, clothes, clock, door, floor, donate, siberia, russia, my, majesty, buy, josef, johann, returnee, returned, good, franz, find, therefore, district; **2**: train, assembly, conditions, under, hungary, looking for, soldiers, government, post, millions, one, years, returnee, graz, women, there, german, applications; **3**: return back, turkish, turkey, sophia, should, return, government, gate, pasha, montenegro, constantinople, border, refugees, emigrants, the same, burgarian, bulgaria, august, amnesty, albanese; **4**: number, economical, united, hungary, states, returning, returnees, austrian, big, societies, early, europe, december, german, budapest, emigration, emigrants, worker, amerika; **5**: return, russia, russian, return, government, prussian, polish, ways, london, home, yesterday, refugees, refugees, bukowina, brasil, putting, emigrants, exception, amnesty; **6**: vienna, wounded, speak, soldiers, ships, austrians, king, no, comrades, home, dear, hans, good, general, one, germans, german, remain, asks, soon; **7:** back, again, comrades, under, transport, stand, returnee, children, return, italian, italy, returnee, home, big, first, reception, germany, german, bosnian; **8**: time, should, already, russia, now, still, message, months, man, lemberg, life, captivity, prisoner of war, can, no, returnee, captivity, fees, woman, parents ('refugees') and '*Pascha*' are related to each other. Deeper historical knowledge and an insight into the corpus are needed to recognise that the Serbo-Turkish war from 1876 to 1878 could be at the centre of this topic. At the same time, topic 3 could also depict a very different historical event that took place in 1911, 34 years later: the return of Albanian Malissors, mountain tribes, from Montenegro to Turkey.
- Topic 5, by contrast, is a good example for a misleading topic. Since '*polnisch*' ('Polish') and '*Flüchtlinge*' ('refugees') are the highest rated terms, the topic seems to be about Polish refugees. But a closer look into the corpus has led to the findings that this topic refers to two very different events: the return of Polish emigrants and the return of refugees from Bukowina. Both events, at least for the human observer, have nothing in common.

On the other hand, important topics, such as the return of Galician refugees during the First World War, are missing. This problem can be solved by increasing the number of topics. When generating twelve topics instead of eight, the discourse on Galician refugees appears. The question of how many topics should be generated depends on the research question and on the corpus itself.

STM delivered some impressive results. It is definitely a helpful tool to discover topics. However, just like most of these tools, it lacks the possibility to link the topics to the original text (that is, where the topics occur in the corpus). This context is often the only way to find out why words were selected for a topic. Therefore, the interpretation of the topics remains a real challenge. The same is true if users want to make sure topics are formed from the correctly connected words. A closer look at the corpus in connection with the topics selected by STM has further shown that various topics can be interwoven. Sometimes this makes sense, but other times it does not.

The last but most important conclusion concerns the epistemological value for the research project. Even though the interpretation of the topics turned out to be difficult, the tool pointed out the most frequent topics in the corpus. Since topic modelling was used with the intention of finding important entry points for further (quantitative or qualitative) investigation, this research aim was achieved. A disadvantage, however, is the lack of user-friendliness – it is difficult to use the R package without any programming skills. If these tools are intended for use by Digital Humanities scholars and especially by a broader audience, this should be high on the agenda of computer scientists and interface designers.

2.4.2 Creating topics on woman's suffrage using TidySupervise

Contrary to STM, TidySupervise relies on supervised classification. The model is built upon a manually labelled training corpus and attempts to identify the words that are regularly associated with the labels. While unsupervised models have to be interpreted at the end of the classification, here the interpretive work comes first. Supervised models consequently offer less flexibility: They take no new information into account and only apply the labels and the corpora that have already been predefined.

While supervised models require significant manual input before being ready for use, there nevertheless is a significant advantage: The results do not need to be reconstructed *ex post*. The models are fixed and transferable across corpora and the processing is much quicker: It takes only minutes to classify one entire month of newspaper issues, and several hours to deal with decades. Unsupervised topic modelling on such a scale would usually take days. Recent studies in Cultural Analytics, e.g. by [Underwood, 2016], have also shown that supervised models can contribute to the analysis of complex cultural phenomena, including the process of genericity or hybridation across genres, through the use of combined probabilities per texts.

For this case study, the 1920–1940 newspaper model of *Numapresse* was reused and trained on 25 issues of four dailies published from 1920 to 1940. Results may be slightly anachronistic for the articles published before 1920 (for instance by stating that some early 1900s texts belong to the movie section, which actually only appeared in 1913 in the French press).

Supervised genre classification with TidySupervise (figure 12) turned out to complement STM-reconstructed topics very well. Some cross-classifications fully converge, such as policy processes with *'institutions politiques'* ('political institutions'), the genre recording the inner workings of parliament and similar institutions (figure 13).

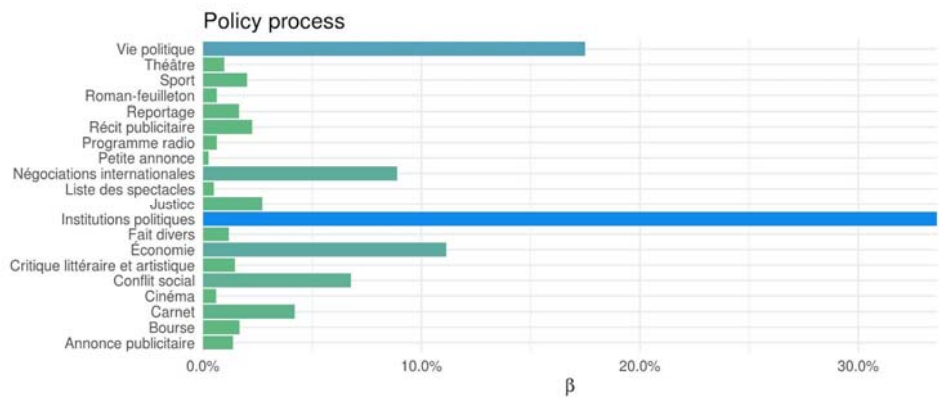

Figure 12. Supervised classifications from TidySupervise on policy processes.

Other alignments are not perfect, but fairly understandable. The feminist movement is both linked to the political news (*'vie politique'* and *'institutions politique'*), but also to social conflict news (*'conflit social'*), as figure 13 shows. This actually echoes the main feature of the movement, which pushed forward a political agenda to emancipate half of the population from gender discrimination.

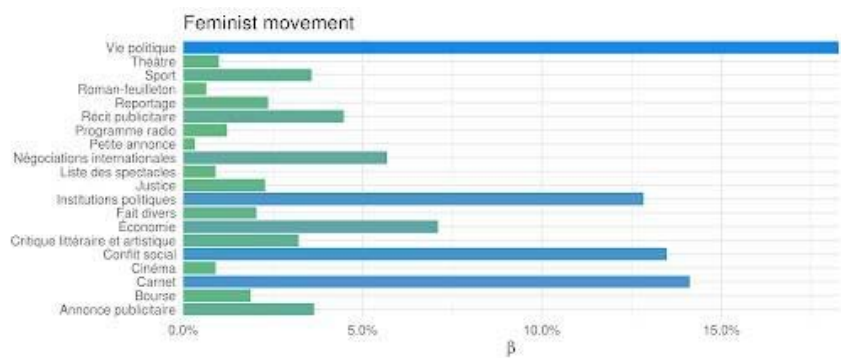

Figure 13. The supervised classifications from TidySupervise on the feminist movement.

Finally, in figure 14, advertisements turn out to be frequently classified as 'announcements', hence the spread of the *'carnet'* (register) genre. This is not really a wrong classification, since notices of feminist meetings are not necessarily paid ads. It may also be partly a consequence of the limitations of Optical Layout Segmentation within the Europeana Newspaper project. Ads and announcements are generally regrouped in wider sections and there is no way to identify the precise ads using the expression *'vote des femmes'*.

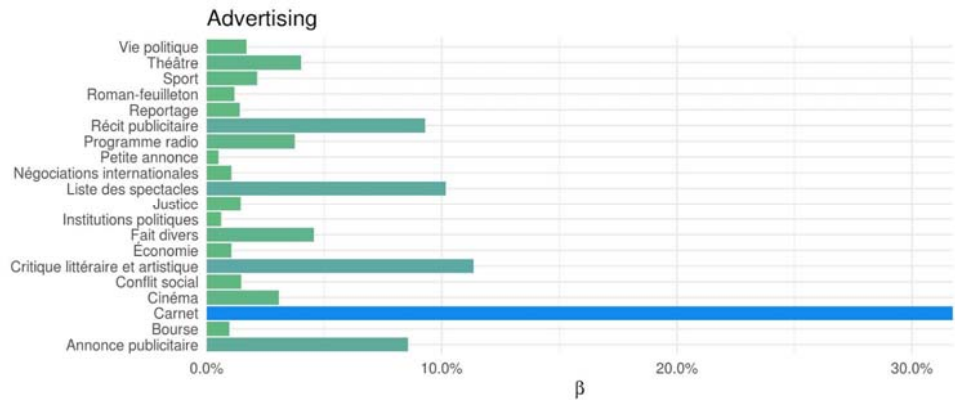

Figure 14. The supervised classifications from TidySupervise on the feminist movement in connection with advertising.

Supervised classification can also be useful to detect hapax legomena or anomalies, while unsupervised topic modelling is more focused on wider classification. In the results, there are a handful of articles classified with a high probability as serial novels. This instance could still be an important signal of wider acculturation processes of the issue of women's suffrage, beyond the political column. Such a hapax would never have been singled out using only an unsupervised model, as classifications are usually derived from larger clusters of documents. Since it is drawn from an external corpus, supervised classification can highlight such unusual outputs.

2.3.1 Creating topics on return migration using Overview

When it comes to topic modelling, Overview explores word patterns using a rather different process than other tools. The focus lies on the comparison of two documents in order to see their similarity. This comparison is done by multiplying the frequencies of equal words and then adding up the results. Documents with high similarities are grouped together using a clustering algorithm based on this similarity of scores. Overview then categorises the documents into folders, sub-folders, sub-sub-folders, and so on (figure 15):

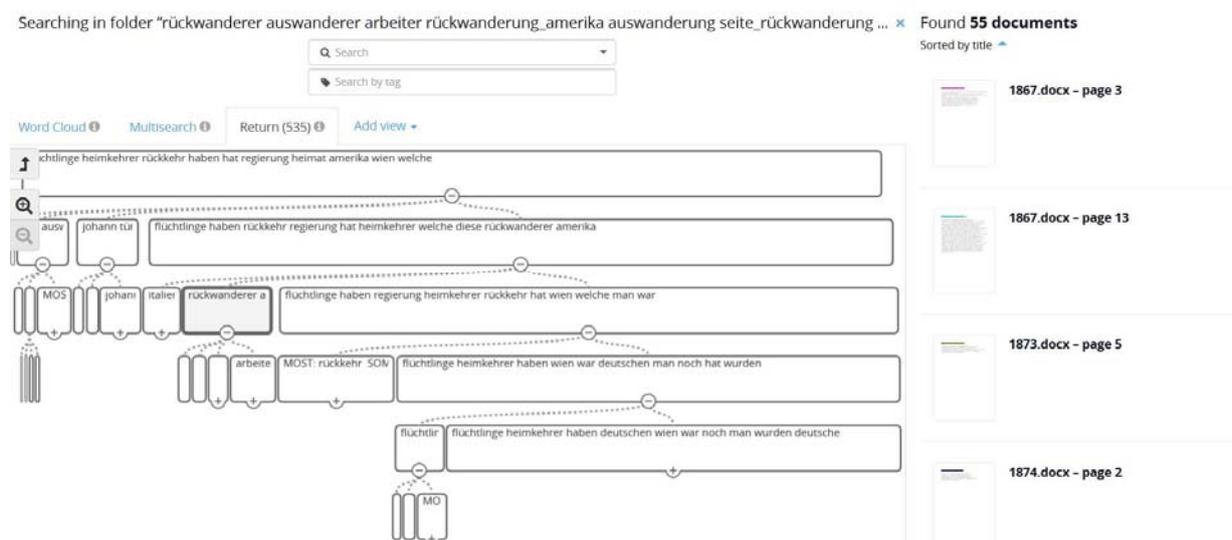

Figure 15. Folders created with Overview.

Overview was used to find topics in the subcorpus on return migration. For this purpose, 77 Word documents (each document covering one year, each page an article) were uploaded to the open source platform. The program removes stopwords automatically (German can be chosen as a language). In addition, it is also possible to remove irrelevant words manually (such as recurring titles of the newspapers etc.). The program then compares every pair of documents (535 pages in total). Just like STM, Overview recognised some of the main themes, such as the Serbo-Turkish War from 1876 to 1878, or the return from America. The topic about the return from America, for example, was found in 55 articles and the folder was labelled with words such as *'Rückwanderer'* ('returnees'), *'Auswanderer'* ('emigrants'), *'Arbeiter'* ('workers'), *'Rückwanderung'* ('return migration') and *'Amerika'* ('America'). The same folder was split into several sub-folders with even more specific topics (each folder splits into smaller and smaller sub-folders, each of which contains a smaller number of documents). One of the sub-folders, for example, was labelled with the words above, but also with *'Hamburg'*, *'Krise'* ('crisis'), *'Dezember'* ('December'), *'allerhöchste'* ('at the very most'), *'Bremen'* and *'europäischen'* ('European'). This topic occurs in twelve articles and maps the news coverage in the subcorpus on remigration from America as a consequence of crises.

Because of the many sub-folders, even more hidden topics can be revealed. For example, six articles were found talking about care and help, particularly for returned emigrants. In another sub-folder, requests and small advertisements from and for returnees were correctly put in relation to each other based words such as *'Heimkehrer'* ('returnee'), *'Posten'* ('post/position'), *'verheiratet'* ('married'), *'bittet'* ('asks/begs') or *'Anträge'* ('requests'). Even though most of the generated topics were understandable, sometimes no obvious common topic could be recognised. In this instance, it was helpful to have the original text at hand, a feature implemented in Overview.

All in all, Overview seems to have some advantages over other topic modelling tools:
- The platform is very user-friendly and easy to use.
- Interactive topic modelling: The tool allows users to iteratively refine the topics by adding and excluding keywords.
- Interactive visualisation of topic model: The visualisation can be 'unfolded', which allows the user go a step deeper with every unfolding.
- The tool displays where the topics occur in the corpus: This makes the results comprehensible and the topics understandable.
- Overview offers a personal workspace: This allows users to annotate (tag) the documents and to save the results.

The overall conclusions that can be drawn from the extensive testing of topic modelling tools are that many of the available tools are not flexible enough for what an increasing number of humanities researchers and a general audience need. To defer these challenges to (often wrongly assumed) users' inability to grasp complex issues is far too simple. Users of digital (newspaper) interfaces need to be able to receive good results without computational skills, detailed knowledge about data structure and management or the basic functionality of topic models. As it is, results often cannot be interpreted by users, because the computed topics do not make sense to the human eye. Interpreting clusters of words can therefore be tiresome and frustrating. Ultimately, the feedback gathered for this paper also underlines [Yang's et al., 2011] conclusions:

> 'We have found that we can automatically generate topics that are generally good, however we found that once we generated a set of topics, we cannot decide if it is mundane or interesting without an expert.[. . . ] We have come to the conclusion that it is essential that an expert in the field contextualise these topics and evaluate them for relevancy.'

We also found that topic modelling seems to be promising for text classification such as genre classification [Langlais, 2019] or for theme classification, and can be an important approach in order to remove false positives (e.g. when identifying irrelevant articles for the research question) from a search result. For other challenges experimented with (but which we cannot address here), word embeddings showed encouraging results for finding related keywords, related articles for a specific query or when exploring language change over time. [Pivovarova, Marjanen and Zosa, 2019] for example used diachronic word embeddings of several timeslices to study the ideological terms ending with an *-ism* suffix. Also, [Tahmasebi, 2018] detected word sense changes using word embeddings on Swedish historical newspapers. In addition, the combination of topic modelling and word embedding opens up new possibilities for research in historical newspapers.

**III CONCLUSION**

The intensive testing of the existing newspaper search interfaces ANNO, Digi and Korp, as well as Gallica and Retronews, showed both similarities and some differences, all of which influence the search outcomes in different ways. ANNO, for example, has the largest collection and, like Digi, offers a very good distance search option. Digi allows users to download the entire collection for further processing, Gallica gives very useful informative historical introductions for topics that have been chosen by the designers of the interface, and Korp provides morphologically analysed text which helps with linguistically oriented research. While both the National Libraries of Finland and of France have separate search interfaces for advanced analysis, the Austrian National Library only recently launched a similar page (ONB Labs) and cannot yet offer a broader service. Then again, while access to all interfaces is free in Finland and in Austria, in France users have to pay for access and use of Retronews, the advanced analysis tool of BNF.

Therefore, although the starting position is different for all interfaces and the outcomes vary accordingly, some overarching conclusion can be made. However, it has to be stressed that not all of the items mentioned below are issues on all interfaces, on the contrary – the websites designed for specialised analysis, Korp and Retronews, already have some of the desired features implemented.

1. In some cases, the OCR quality is still very low. After identifying some major issues in this regard, the DH team's reliance on (and trust in) some search results was very low.
2. Both the surveys and the NewsEye testing showed that keywords and context are key for success: Topic and keyword suggestions could help users to achieve better results. Improvement of the search options (e.g. Boolean operators, asterisks, wildcards and regular expressions) would help advanced users in particular. The use of regular expressions is often limited because of users' fear of creating queries that require lots of computing. It might be better to limit searches in another way than setting restrictions for researchers' creativity.
3. For some research cases (such as genealogy, the history of places, cities or organisations, associations and clubs, or historical network analysis, etc.) tools that use named entity recognition (and possibly linking) would be especially important, particularly for family historians and chroniclers, as well as for many academic analysis and visualisations. Optical layout recognition, like article separation, image or advertisement recognition, on the other hand, would allow for a targeted and complex analysis of easier to select corpora.
4. The possibility of a personal workspace with functions to create subcorpora and annotation tools also gained the respondents' and the NewsEye DH teams' interest. A personal workspace that includes possibilities both to include and to exclude search results, to limit corpora, to manually annotate and to use advanced functions for analysis and visualisations, download options for newspapers, articles or entire corpora and metadata were asked for in connection with the need for quantitative and qualitative analysis.
5. Tools for analysis and visualisation both on the interfaces and in a personal workspace would greatly improve the understanding of performed searches and help to contextualise results, as the DH team of NewsEye was able to show. Among the tools requested and addressed were frequency analysis, KWIC analysis (keyword in context), co-occurrences, context visualisation, word clouds, different graphs, publication periods, amount of issues, and geographical visualisations. Simply put, text mining options would be welcomed by newspaper researchers. This is also true for metadata: Number of words, pages, languages etc. per year/month/day are needed for in-depth analysis and interpretation.
6. Although libraries might struggle to offer appropriate functions for quantitative analysis and visualisations on their newspaper interfaces, it has to be taken into consideration that these can point to errors in the dataset, such as bad OCR quality, missing data, incorrect metadata, etc. To implement such a tool could therefore ultimately be to the advantage of the libraries.
7. Frequencies are needed if topic modelling approaches are considered as alternative ways to analyse the datasets. The DH teams of NewsEye tested various topic modelling tools, some of them implemented on the newspaper interfaces, but most of them available on other sites. It can be concluded that despite the ongoing issues with the available tools, topic modelling could be advantageous when dealing with the mass of data that is available on newspaper interfaces.
8. It is not enough to create powerful tools: Their accessibility should be guaranteed by the mediation of professionals (e.g. librarians). Many different user groups are rather proficient at using the interfaces, but there is ample opportunity for improvements.

As an overall assessment, it can be underlined that digital tools combining quantitative macro-analysis (big data analysis) and qualitative microanalysis (reading) have some significant advantages over other methods. Especially when dealing with massive amounts of data, the sheer quantity makes the traditional practice of reading in context untenable and an inadequate method of evidence gathering. Analysing this data with traditional micro-analysis approaches often turns out to be impossible, and researchers get the impression that they are missing out on important things. At the same time, the arguments vouching for quantitative approaches can be used against macro-analysis: By looking at the data from afar, crucial things could be missed. Agreeing with [Jockers, 2013], it must be said that the two ways of analysis, therefore, should and must co-exist. The digital newspaper interfaces studied for this paper, however, do not yet offer opportunities for both approaches. The tools and methods developed within the NewsEye project could therefore highlight the chances and challenges of complex digital interfaces.